\documentclass[10pt,journal,compsoc]{IEEEtran}
\ifCLASSOPTIONcompsoc
  \usepackage[nocompress]{cite}
\else
  \usepackage{cite}
\fi

\usepackage[pdftex]{graphicx}
\graphicspath{{../}{../}}
\DeclareGraphicsExtensions{.pdf,.png}

\usepackage{algorithm}
\usepackage{algpseudocode}
\usepackage{adjustbox}

\usepackage[caption=false,font=footnotesize,labelfont=sf,textfont=sf]{subfig}

\begin{document}
\title{MultiViz : A Gephi Plugin for Scalable Visualization of Multi-Layer Networks}

\author{Jayamohan~Pillai C.S,~
        Ayan~Chatterjee,~
        Geetha~M,
        and~Amitava~Mukherjee~
        
\IEEEcompsocitemizethanks{\IEEEcompsocthanksitem Jayamohan Pillai (\textit{Corresponding author}) and Geetha M are with the Department
of Computer Science, Amrita School of Computing, Amrita Vishwa Vidyapeetham, Amritapuri, Kerala, India,\protect\\
E-mail: amenp2cse20006@am.students.amrita.edu\\(kanha.jay@gmail.com),  geetham@am.amrita.edu,
\IEEEcompsocthanksitem Ayan Chatterjee is with Network Science Institute, Northeastern University, Boston, MA, USA.\\
E-mail: chatterjee.ay@northeastern.edu
\IEEEcompsocthanksitem Amitava Mukherjee is withAmrita  School of Computing, Amrita Vishwa Vidyapeetham, Amritapuri, India and College of Nanoscale Science and Engineering, SUNY Polytechnic Institute, Albany, NY.\\
Email: amitavamukherjee@am.amrita.edu
}}

\IEEEtitleabstractindextext{%
\begin{abstract}
The process of visually presenting networks is an effective way to understand entity relationships within the networks since it reveals the overall structure and topology of the network. Real networks are extremely difficult to visualize due to their immense complexity, which includes vast amounts of data, several types of interactions, various subsystems and several levels of connectivity as well as changes over time. This paper introduces the "MultiViz Plugin," a plugin for gephi, an open-source software tool for graph visualization and modification, in order to to visualize complex networks in a multi-layer manner. A collection of settings are availabe through the plugin to transform an existing network into a multi-layered network. The plugin supports several layout algorithms and lets user to choose which property of the network to be used as the layer. The goal of the study is to give the user complete control over how the network is visualized in a multi-layer fashion. We demonstrate the ability of the plugin to visualize multi-layer data using a real-life complex multi-layer datasets. The plugin is open-source and is freely available at https://github.com/JSiv/MultiViz.
%https://github.com/ChatterjeeAyan/MultiViz
\end{abstract}

\begin{IEEEkeywords}
Multilayer Networks, Gephi Plugin, Graph Visualization, Network Visualization
\end{IEEEkeywords}}

\maketitle

\IEEEdisplaynontitleabstractindextext

\IEEEpeerreviewmaketitle

\IEEEraisesectionheading{\section{Introduction}\label{sec:introduction}}

\IEEEPARstart{U} {nderstanding} complex networks enable us to understand a wide range of real systems, ranging from technological to biological networks systems. Protein interaction networks, transportation networks, and social networks are all examples of complex networks. Initially, network studies treated networks as conventional graphs, with nodes (vertices) representing entities or agents and edges acting as a link between nodes. When exploring more realistic and complex systems, however, many factors must be considered, such as directed graphs\cite{barabasi1999emergence}\cite{clauset2009power} \cite{bang2008digraphs}, weighted graphs\cite{newman2004analysis}\cite{barrat2004architecture} \cite{clauset2009power}, bipartite graphs \cite{barabasi1999emergence} \cite{clauset2009power} \cite{breiger1974duality}, network of networks \cite{d2014networks}, and so on.

Different forms of complex network graphs with multiple subsystems can be found in the domains of life sciences, sociology, digital humanities and many other fields \cite{mcgee2019state}. \textit{Interdependent networks} in which nodes from two or more basic networks are connected via "dependency edges," with any failure in the dependency edges potentially affecting all networks connected by it. \textit{Interconnected networks} and \textit{networks of networks} where nodes from different networks are adjacent to one another, but are not dependent on each other. \textit{Multitype networks} \cite{cai2005community}  with nodes of the same or different "type" adjacent to each other.\textit{K-partite networks} \cite{barabasi1999emergence} \cite{clauset2009power}, in which nodes of the same colour or type cannot be connected by an edge. \textit{Multiplex networks} \cite{nicosia2013growing} \cite{verbrugge1979multiplexity}  \cite{bianconi2013statistical} \cite{cellai2013percolation} are complex networks containing various types of edges, with each sub network sharing at least one node with the others. \textit{Hypergraphs} \cite{berge1984hypergraphs} are a type of complex network in which any number of nodes can be connected to any number of edges.

Most of these complex networks can be fashioned into a multi-layer network by considering another dimension called layers, along with the standard graph $(G = {V, E})$ representation of the network, using nodes which represent an entity or an agent and edges that connect a pair of nodes. Multi-layer visualization $(G = {V, E, L})$ can be explained as a multi-dimensional representation of complex networks, where $V$ is the number of vertices, $E$ is the edges that connect a pair of vertices and $L$ denotes a set of layers. Each layer can include any subset of nodes, with nodes in one layer connecting to nodes in another layer (inter-layer edges) and nodes in the same layer connecting to each other (intra-layer edges) \cite{kivela2014multilayer}. We can generalize multi-layer networks in to two types, layer-disjoint networks and node-aligned networks\cite{kivela2014multilayer}. Layer disjoint networks feature different set of nodes in each layer, and node-aligned networks can have the same nodes in different layers.

There are numerous applications available for visualizing networks. Gephi is one of the major generic large-scale network visualization and analysis tools out there \cite{pavlopoulos2017empirical}. But despite being the most cited and widely used \cite{jacomy2019gephiicwsm}, gephi lacks a proper multi-layer visualization plugin.

The proposed method is based on an existing plugin named Network Splitter 3D \cite{barao2014network}. The Network Splitter plugin helps to vertically separate a network into Z levels/user-defined clusters. We adapt and enhance this technique to visualize a multi-layer network by adding various functionalities to the plugin.

\section{Related Works}
Visualization is a critical aspect that provides humans with a significant edge in identifying useful features in data or a network. We live in a complex world where everything is connected. Large scale data-sets are abounded in every field of human activity, from biology to medicine, economics and climate research. These datasets portray massive real-world complex networks. These networks are rich in obscured domain-specific data which can be extracted and studied using multi-layer visualization. 

Some of the major tools for multi-layer network visualization make use of python and R-language libraries. NetworkX \cite{hagberg2008exploring}, a python module for creating, manipulating, and studying the structure, dynamics, and functions of complex networks. and Matplotlib \cite{Hunter:2007}, a python package that allows you to create static, animated, and interactive visualizations are employed by several applications such as MultinetX\cite{nkoub2015multi} and Py3Plex \cite{vskrlj2018py3plex} to display multi-layer networks. These applications expect some amount of programming knowledge from the user. R-language users usually prefer the igraph library \cite{csardi2006igraph} for network visualization. Muxviz\cite{de2015muxviz} is a multi-layer network visualization and analysis tool-kit for the R environment. It enables GUI-based analysis solutions, unlike networkX-based tools. Arena3D \cite{pavlopoulos2008arena3d} focuses on visualizing multi-layer networks in 3D.

In this big data era, we are dealing with massive networks with various features and topologies with millions of nodes and connections. The activities of visualization and exploration have become extremely computationally demanding. Gephi \cite{bastian2009gephi}, Cytoscape \cite{shannon2003cytoscape} \cite{smoot2011cytoscape}, and Pajek \cite{batagelj2004pajek} are currently the strongest candidates for large-scale network visualization.

Cytoscape \cite{smoot2011cytoscape} is the most extensively used 2D network visualisation tool in biology and health sciences. But its layouts are highly memory and CPU hungry and it can't handle enormous amounts of data. Pajek \cite{batagelj2004pajek}, a Microsoft Windows-based network visualisation program that outperforms every other tool in the field in visualising millions of nodes with billions of connections in an average computer but it only accepts a few file formats and its lack of operating system interoperability, input file format flexibility, and appealing visualisations prevent it from being used for complex visualisations. Gephi is a free open-source network visualization and exploration application and a better candidate than its counterparts when coming to visualization of generic large-scale networks\cite{pavlopoulos2017empirical}. It is more memory efficient, user friendly and stable than Cytoscape and it also supports a wide variety of input file types such as GEXF, GDF, GML, GraphML \cite{brandes2013graph}, Pajek (NET), GraphViz (DOT)\cite{ellson2001graphviz}, CSV, UCINET (DL), Tulip (TPL)\cite{auber2017tulip}, Netdraw (VNA), and Excel spreadsheets. In this paper, we propose to visualize multilayer complex networks using the Gephi Application.

\section{MultiViz : Multi-layer Network Visualization}
Visualizing complex networks in a multi-layer fashion helps in identifying the structures hidden in the network. The network can be partitioned into layers based on their properties, each layer can be a community, a sub-network or a group of nodes. Even multiple networks can be compared with each other by considering each network as a layer. For instance, by comparing a person's various social media accounts such as Facebook and Twitter, we can spot trends in the types of people or topics they connect with and provide answers to questions like: Are they more likely to connect with their friends and co-workers? Or are they more interested in politics or celebrity news? Or do they connect with the same "type" of people or topic across all of their social media accounts?, Or do they use each account for a different purposes? In this scenario, each social media account may be thought of as a layer, with nodes and edges that can be coloured or sized to highlight various features in the dataset. Communities can be detected in each layer, and nodes can be coloured or categorised in accordance with the communities that have been identified. By visualising this network, we can spot trends in each layer, aiding in further network research. 

Complex real world networks have many features and dimensions, but not all of them can be represented by a conventional graph. Hidden dimensions of a complex network can be visualized through layers. Consider a network for air travel where the nodes are the airports and the edges are the routes that connect each airport; the size of the node depends on the number of routes that pass through the airport; the position of the nodes can be determined by the location of each airport; the colour of the nodes can be determined by the country to which the airport belongs; and the colour of the edges can be determined by another aspect or feature of the dataset. However, the dataset may also contain a wide range of other information, such as all the planes that pass through a specific airport and different types of aeroplanes, among other things. We can utilise layers to display these additional network dimensions. Each layer could represent a specific type of aircraft (e.g., jumbo jets, mid-size jets, etc.) or airline ( eg: Air India, Japan Airlines, United Airlines etc). Some features like node colour, position etc will be similar in both layers, Other features like node size will change as each layer now shows only routes limited to an airline or aeroplane type. Therefore, by splitting the network we can compare different flight paths or aeroplane routes. Users can identify similar trends in different layers, connections between two layers and so on when visualizing network in a multilayer fashion.

Most real-world complicated networks are enormous, necessitating the need for a special tool to visualize them. We are utilising the Gephi application to visualize multi-layer networks in this research. Users can import and generate networks in Gephi. Our multi-layer visualization plugin then allows them to perform several visualization algorithms (force atlas2\cite{jacomy2014forceatlas2}, Fruchterman reingold\cite{fruchterman1991graph}, and so on...) in a multi-layer fashion. In this section, we showcase the MultiViz plugin and its implementation.

\section{Methodology and Implementation}
The visualization approach using multiviz plugin is aimed at resolving the problem of displaying complex real-world networks with multiple layers of connectedness. The presented layout algorithm implementation is given in detail below.

\subsection{User Options}
Users can either import a network into gephi or they can create their own graph. The data laboratory of gephi provides the user with a node table and an edge table which allows users to add, delete, duplicate and merge columns in the tables and they can also add, edit and delete nodes and edges in the network.

\begin{figure}[!t]
\centering
\includegraphics[width=0.68\linewidth, scale=10]{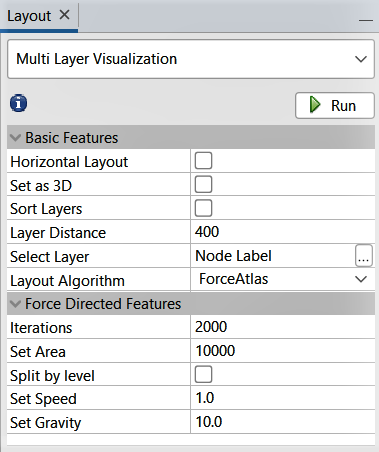}
\caption{MultiViz plugin for Gephi. The \textit{Select Layer} option lets user choose the network feature which is to be considered as the layer. User can control the distance between two layers, number of iterations, speed, area and gravity for force directed layouts. User can also choose which layout algorithm to apply in each layer. After providing required values in the layout plugin user clicks the '\textit{Run}' button, by which the multiviz plugin create a multilayer visualization of the network which they have loaded in gephi}
\label{fig:plugin}
\end{figure}

Multiviz plugin (Figure~\ref{fig:plugin}) allows users to choose the distance between two layers, the graph feature which is to be used as a layer, whether to display the graph vertically or horizontally and the number of iterations, speed and gravity for forced directed layouts. The plugin uses either force directed layouts (Fruchterman-reingold, forceatlas) or basic layouts (circle, grid and linear layouts) to visualize the network.Option to sort the layers based on the number of nodes in each layer is also possible with the plugin. The split-by-level option lets user run the selected layout algorithm on the whole network and then split the network into layers thereby keeping the positions on the node similar to that of the underlying network. This option works similar to Network Splitter 3D\cite{barao2014network} plugin in gephi. If the split-by-level option is not selected, the selected layout algorithm will be applied on each layer. When level splitter is selected the output network can be used to view the exploded view\cite{wikiview} of the network. The multilayer network visualized without level splitter can be used to compare multiple layers as networks to identify similar structures, trends and patterns.

\subsection{Layer Splitting Algorithm}
The multiviz plugin initially assigns each node to a layer, based on the user input. The user-selected layout algorithm is then applied on the first layer. When the node placements on the first layer is completed, the farthest node on the layer is selected and the following layer is placed at a constant distance from that selected node. The distance between two layers can be edited by the user. By default, the multilayer network is visualized in a vertical manner by stacking one layer on top of another. The user can option to view the network in a horizontal plane by selecting the option in the plugin.

\begin{algorithm}[H]
\caption{Splitting network into layers}
\begin{algorithmic}
\Require $nodes, edges \gets $nodes in the network and edges between them
\Require $isHorizontal \gets $ horizontal or vertical layer placement
\Require $is3D \gets $ 3D or 2D layer visualization
\State $selectedLayer \gets$ user selected feature as layer
\State $layerDistance \gets$  user defined
\State {each node is assigned to the layer it belongs}
\State $layers \gets [node.property == selectedLayer]$
\State $previousLayer = firstLayer$
\For{$currentLayer\ in\ layers$}
\State $farthest \gets$ farthest node from the previous layer
\State $biggest \gets$ biggest node in the current layer
\If{$previousLayer \neq currentLayer$}
\State $fy \gets farthest.y() + farthest.size()$
\State $fy \gets fy + layerDistance$
\Else
\State $fy \gets firstLayer.getnodes[0].y$
\EndIf
\State $area \gets biggest.size() + biggest.label * nodes.size()$
\State $layerNodes \gets currentLayer.getnodes()$
\If {$isHorizontal$}
    \For{$node\ in\ layerNodes$}
        \State $\theta \gets 270 * \pi /180$
        \State $x \gets node.x, y \gets node.y + fy$
        \State $node.x \gets  xCos\theta - ySin\theta $
        \State $node.y \gets yCos\theta + xSin\theta$
    \EndFor
\Else
    \For{$node\ in\ layerNodes$}
        \State $node.y \gets node.y + fy$
    \EndFor
\EndIf
\If {$is3D$}
    \State $\theta \gets 65 * \pi /180$
    \For{$node\ in\ layerNodes$}
        \State $node.y \gets node.y * Cos\theta - node.z * Sin\theta$
        \State $node.z \gets random * 0.01$
    \EndFor
\EndIf
\State $previousLayer = currentLayer$
\EndFor
\end{algorithmic}
\end{algorithm}

When working with force-based layouts, it is possible to create several random force points or layer points to attract the nodes belonging to each layer thereby splitting the network in to multiple layers. Each of these points will act as a placeholder for the layers. It is seen from the experiments that the attractive force in these points should be greater than the default attractive force for the layer split to happen. We haven't integrated this algorithm in the plugin since these new attractive forces greatly affect the shape of the layer and may lead to a biased placement of the nodes.

\subsection{Layout Algorithms}
We have used multiple layout algorithms in our multiviz plugin, including some basic layouts like Circle, Linear, Random and Grid Layout,  and force-directed layouts like Fruchterman Reingold and Force Atlas. When determining the distance between two nodes in the layout algorithms, we took into account the size of the nodes as well as the size of their labels to avoid overlapping.

\paragraph{Basic Layouts}
The \textit{circle layout} is a simple layout which arranges nodes in a circle. The \textit{grid layout} arranges all the nodes in a grid format and the \textit{linear layout} arranges all nodes linearly one after the other in a single line. In the \textit{random layout}, each node is randomly placed inside a layer. The size of the nodes is not taken into account, and the arrangement may result in node overlap. These layouts are available by selecting Layout -$>$ Layout Name. When the sort option is enabled, the layers will be stacked based on the number of nodes within the layer, with layers with less number of nodes placed at the bottom.

\paragraph{Force Directed Layouts}
Multiple force-directed layouts are implemented in this plugin. In the \textit{Fruchterman-Reingold} force-directed layout algorithm, the nodes are considered steel rings, and the edges are the springs that connect them. The repulsive force used is analogous to the electrical force, and the attractive force is comparable to the spring force. The main goal of the algorithm is to lower the system's energy usage by relocating the nodes and altering the forces between them. The sum of the force vectors determines which direction a node should go and when the system's energy is reduced, the nodes cease moving and the system finds equilibrium. The \textit{Force Atlas layout} algorithm is another force-directed algorithm used here. It is a spatial layout algorithm for real-world networks. In this algorithm, the quality of layout is given more importance than the speed with which the layout is computed. \textit{ForceAtlas2} is an updated force atlas layout. It is the default layout algorithm for Gephi. It incorporates several methodologies such as the Barnes Hut simulation, degree-dependent repulsive force, and local and global adaptive temperatures to improve on the force atlas methodology.

\begin{algorithm}[!h]
\caption{Complete Multilayer Random Layout Algorithm}
\begin{algorithmic}
\Require $nodes, edges \gets $nodes in the network and edges between them
\Require $column \gets$ user selected layer
\Require $layer\_distance \gets$ distance between layers
\State $layers \gets [node.property == column]$ nodes are assigned to the layer it belongs to.
\State $previous\_layer \gets first\_layer$
\For{$layer\ in\ layers$}
\State $top \gets$ farthest node in the previous layer
\State $big \gets$ biggest node in the current layer
\If{$layer \neq first\_layer$}
\State {adding some distance between two layers}
\State $y \gets top.y() + top.size() + layer\_distance$
\Else
\State $y\gets n.pos.y$
\EndIf
\State $node.y \gets y$
\State {establishing area of a layer for placement of nodes}
\State $area \gets big.size() + big.label * nodes.size()$
\For{each node $n$}
    \State $x \gets n.pos.x + (-area/2 + a\times random())$
    \State $y \gets n.pos.y + (-area/2 + a\times random())$
    \State $node.set(x,y)$
\EndFor
\State $previous\_layer \gets layer$
\EndFor
\end{algorithmic} \end{algorithm}

\section{Results}
Multiple sorts of relationships between system elements can add complexity to real-world networks. In the field of social network analysis multiple forms of social connection between people can be represented by multiple types of links. To accommodate the presence of more than one type of link, a multi-layer network $G =(V,E,L)$ can be employed, where $L$ is a set of layers (or dimensions), each of which represents a distinct type of link.

In this study, we have developed a plugin tool for the Gephi application, to visualize a given network in a multi-layer fashion to help researchers take advantage of the multi-layer properties of complex networks in their research.

To demonstrate the ability of MultiViz plugin to visualize layered networks, we consider three different multilayer networks. A complex network that shows the types of genetic interactions for Candida Albicans in Biological General Repository for Interaction Datasets (BioGRID \cite{stark2006biogrid}) acquired from manliodedomenico.com\cite{domenico2015candida}\cite{de2015structural}. Candida Albicans is a common pathogen found in human gut flora. BioGrid is a public database that archives and disseminates genetic and protein interaction data from humans and model organisms. The data used is from BioGRID 3.2.108 (updated 1 Jan 2014). Seven layers considered in this network: (a) Additive genetic interaction defined by inequality, (b) Synthetic genetic interaction defined by inequality, (c) Physical association, (d) Direct interaction, (e) Suppressive genetic interaction defined by inequality, (f) Association, (g) Colocalization. The Diseasome \cite{diseasome2019}\cite{goh2007human} Network is a multilayer network that represent genes and symptom relationship that is visualized using multiviz plugin in this paper. Usually the diseases that share genes tend share similar symptoms. Genes associated with similar disorders show both a higher likelihood of physical interactions between their products and higher expression profiling similarity for their transcripts, supporting the existence of distinct disease-specific functional modules. The third network which we have visualized is the London Multiplex Transport Network\cite{de2014navigability}. The data is from 2013 released by the website of transport for London (https://www.tfl.gov.uk/). The nodes is the networks are train stations in London and edges are the existing routes between those stations. Three types of train stations are considered :Underground, Overground and DLR. The datasets used here are acquired from manliodomenico.com.

\clearpage
\begin{figure*}[!h]
\centering
\subfloat[Candida Albicans Genetic Interaction Network]{
	\includegraphics[scale=0.4, trim ={0cm 5cm 0cm 8cm}, clip]{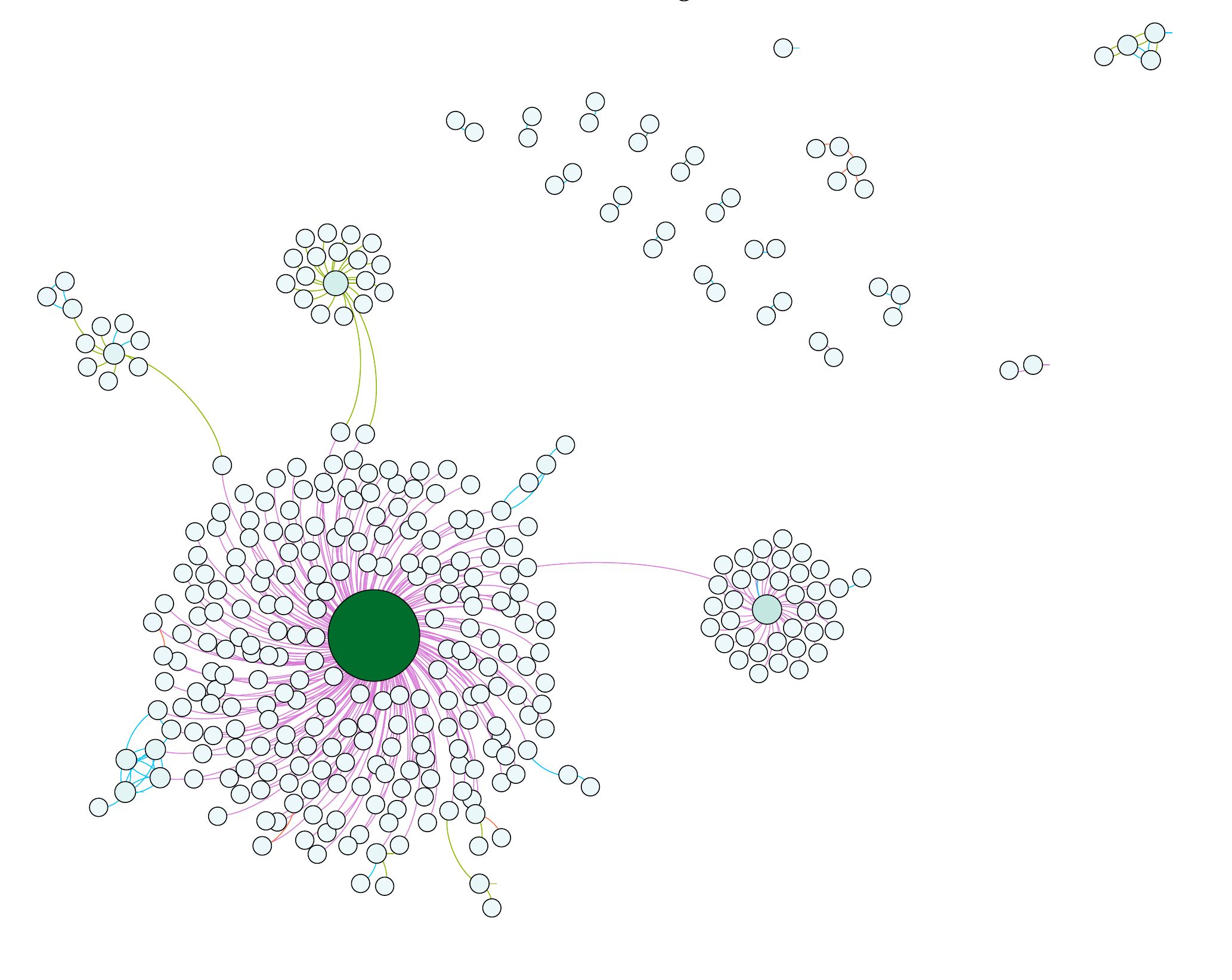}%
	\label{fig:cagin}}
\subfloat[Multilayer Fruchterman-Reingold visualization of Candida Albicans genetic interaction network]{
	\includegraphics[width=0.5\linewidth, trim={0.5cm 5cm 0.5cm 5cm}, clip]{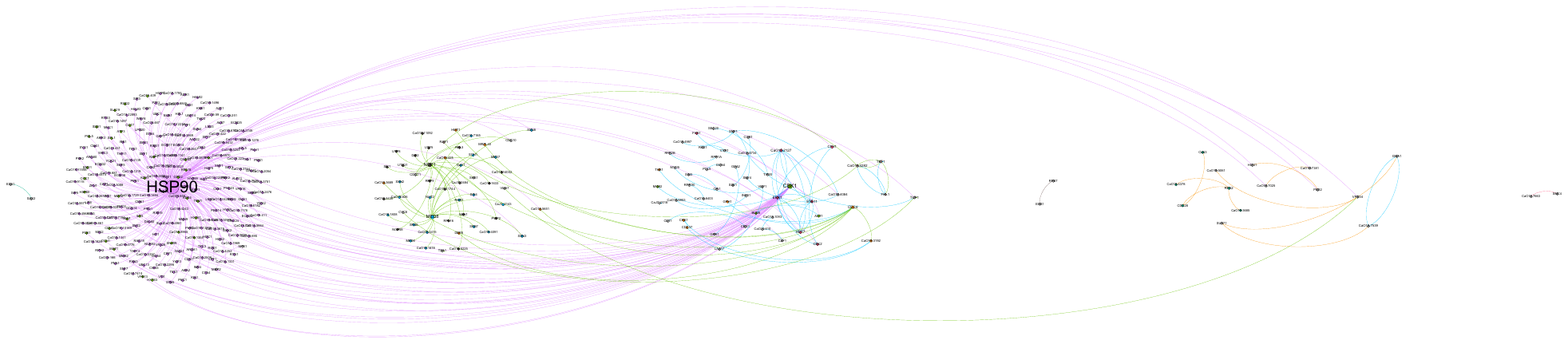}%
	\label{fig:mfr_cagin}}
\hfil
\subfloat[Multilayer ForceAtlas Candida Albicans genetic interaction network.]{
		\includegraphics[width=0.5\linewidth, trim={0.5cm 5cm 0cm 5cm}, clip]{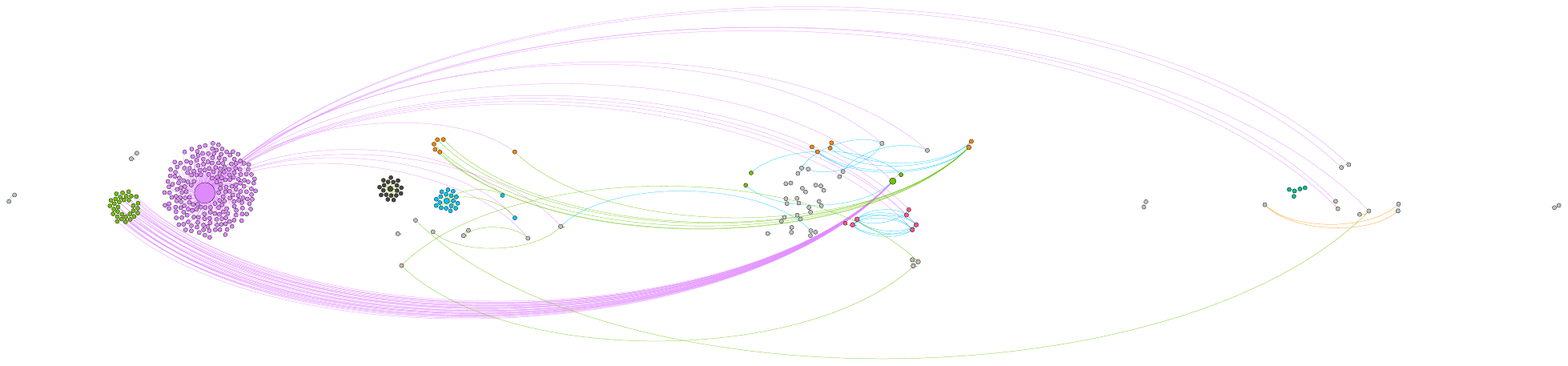}%
		\label{fig:mfa_cagin}}
\subfloat[Multilayer Circular Candida Albicans Genetic Interaction Network]{
	\includegraphics[scale=0.4, trim ={0cm 5cm 0cm 5cm}, clip]{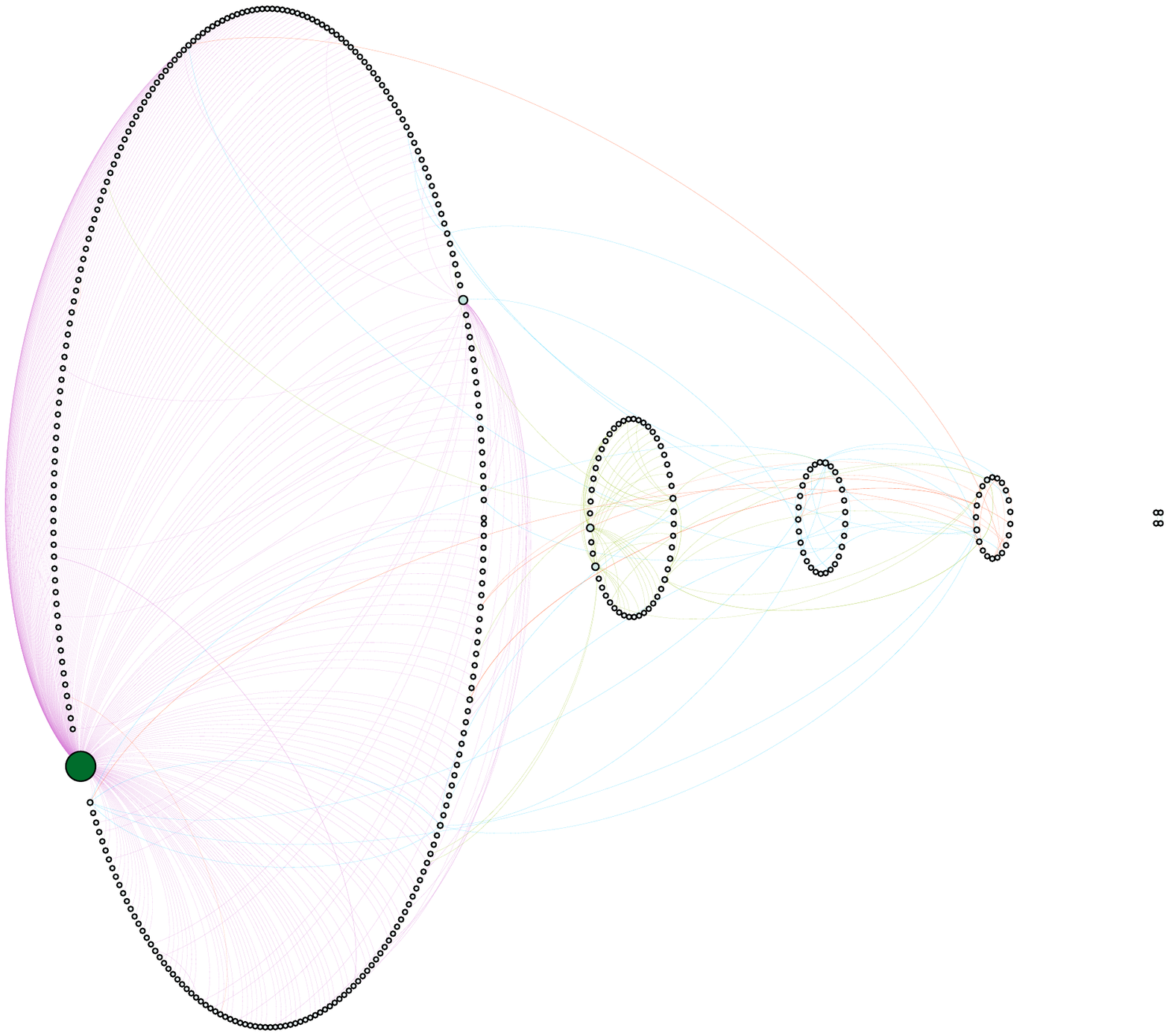}%
	\label{fig:caginx}}
\caption{The above networks, Figure ~\ref{fig:cagin},~\ref{fig:mfr_cagin} and~\ref{fig:mfa_cagin},shows genetic interaction network of Candida Albicans, a most common fungal pathogen. Figure ~\ref{fig:mfr_cagin} and ~\ref{fig:mfa_cagin} shows the candida network in a multilayer fashion, the former used Fruchterman-reingold layout algorithm to arrange nodes and edges in each layer and the latter used forceAtlas layout for the visualization. Each layer can be identified with the color of the edges or links. The size of the nodes depends on the average degree of the node}
\label{fig:cagin_full}
\end{figure*}
  
\begin{figure*}[!h]
\centering
\subfloat[Network of Disorders and Disease Genes]{
\includegraphics[width=0.4\linewidth, trim ={0cm 0cm 0cm 5cm},  clip]{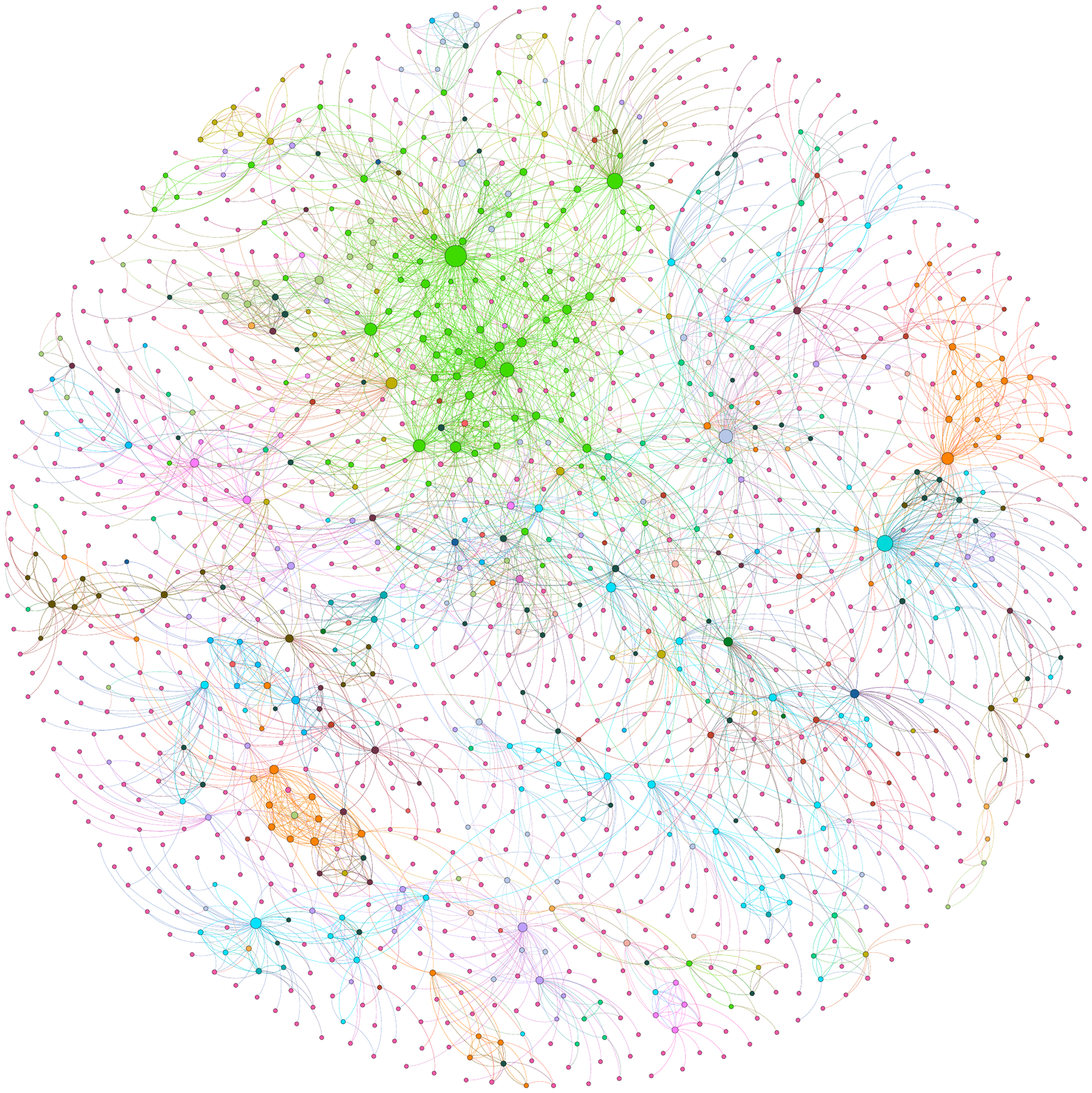}%
\label{fig:ndgor}}
\hfil
\subfloat[Multilayer Fruchterman-Reingold Disorder-Gene Association Network with Level-Split]{
\includegraphics[width=0.4\linewidth, trim={0.5cm 0.5cm 0.5cm 0.5cm}, clip]{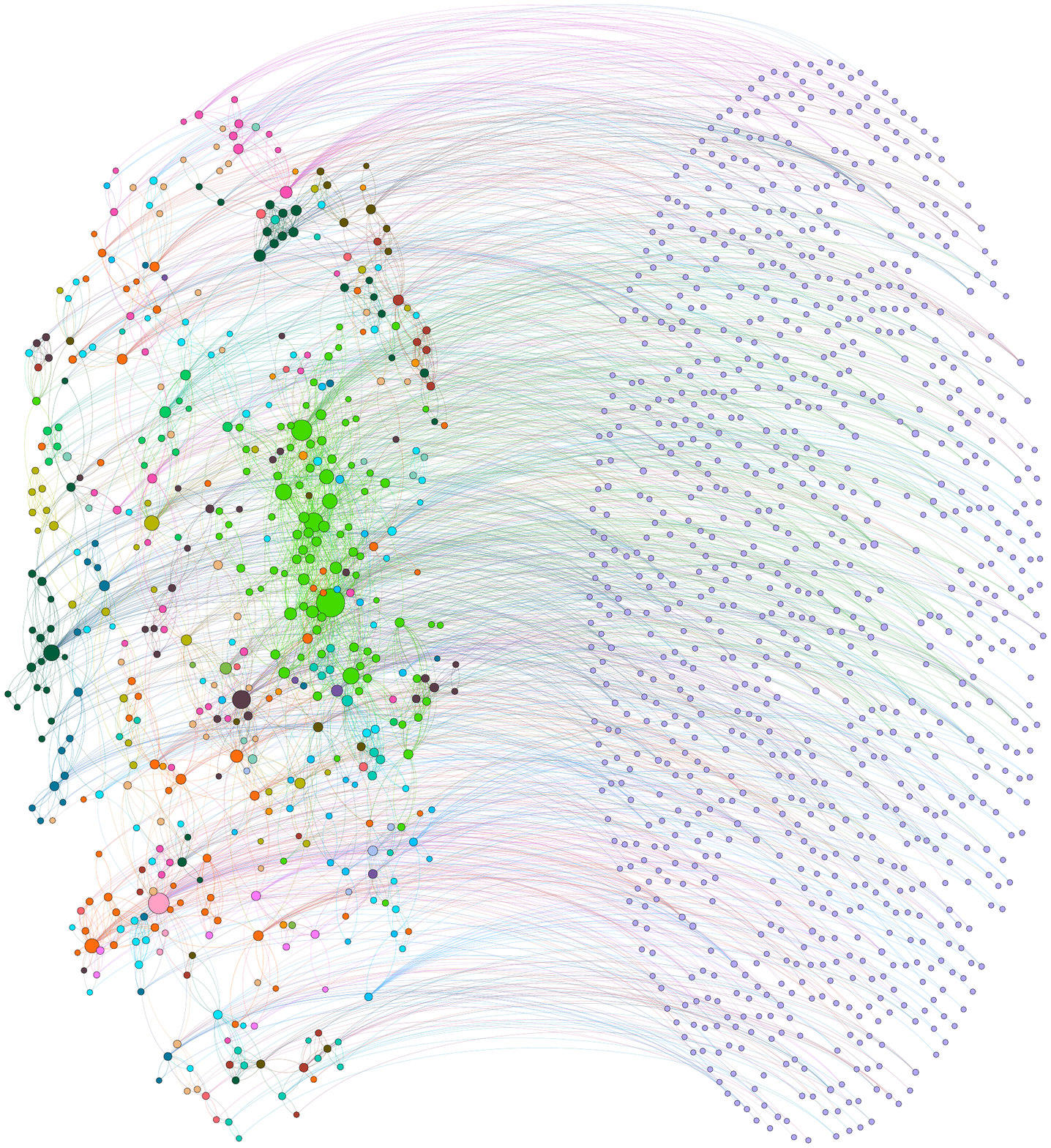}%
\label{fig:fr_ndgls}}
\hfil
\subfloat[Multilayer Fruchterman-Reingold Disorder-Gene Association Network without Level-Split]{
\includegraphics[width=0.4\linewidth, trim={0.5cm 0.5cm 0.5cm 0.5cm}, clip]{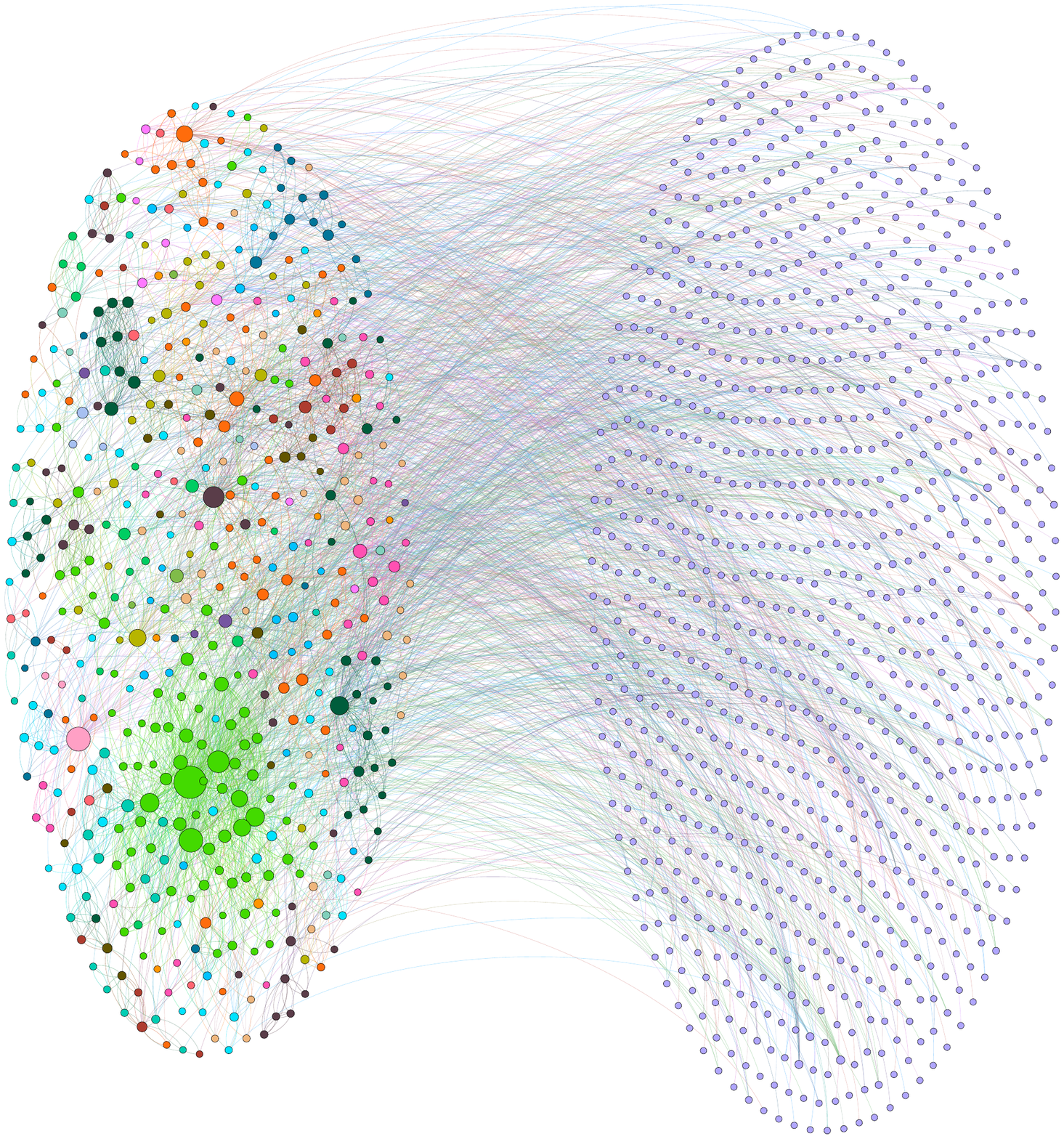}%
\label{fig:fr_ndg}}
\hfil
\subfloat[Multilayer Fruchterman-Reingold Disorder-Gene Association Network (Vertical) with Level-Split ]{\includegraphics[width=0.4\linewidth, trim={0.5cm 0.5cm 0.5cm 0.5}, clip]{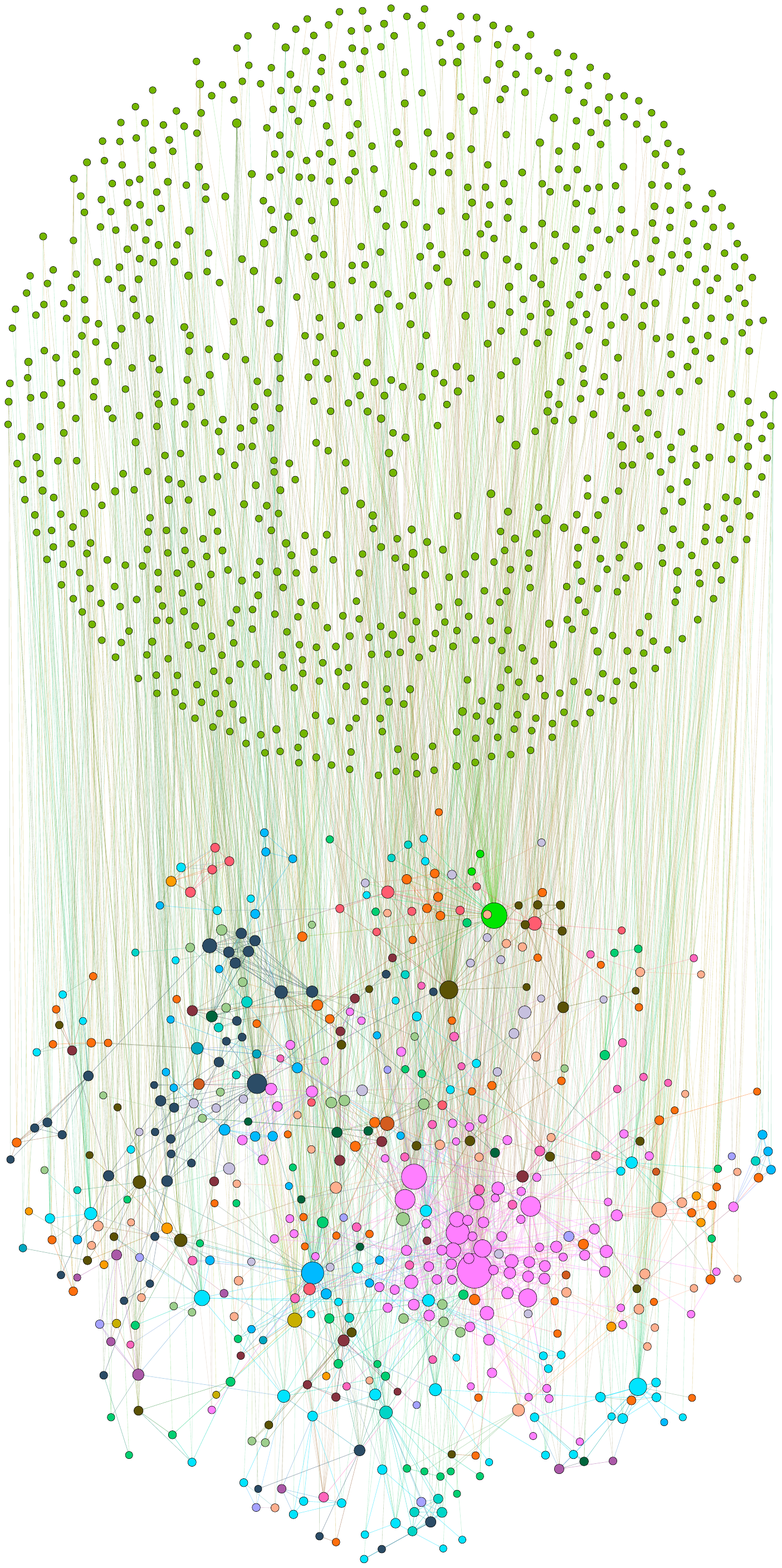}%
\label{fig:fr_ndglv}}
\caption{The above Figures (~\ref{fig:ndgor}, ~\ref{fig:fr_ndgls},~\ref{fig:fr_ndg} and ~\ref{fig:fr_ndglv}) shows Diseasome \cite{goh2007human}, a network of disorders and disease genes linked by the known disorder–gene associations, indicating the common genetic origin of many diseases. Figures(~\ref{fig:fr_ndgls},~\ref{fig:fr_ndg} and ~\ref{fig:fr_ndglv}) shows Diseasome network in a multilayer fashion using Fruchterman reingold layout algorithm. In Figures ~\ref{fig:fr_ndgls} and ~\ref{fig:fr_ndg}, the layers are stacked horizontally where each layer is place one after the other and each layer is displayed in a 3D plane. In Figure~\ref{fig:fr_ndglv}, layers are stacked on top of each other. The color of nodes shows the different types of disorders. The first layer shows all disorders and the second layer shows the corresponding disease genes}
\label{fig:ndg_full}
\end{figure*}

\begin{figure*}[!h]
\centering
\subfloat[London Transportation Network]{\includegraphics[width=0.4\linewidth, trim={0.5cm 0.5cm 0.5cm 0.5cm}, clip]{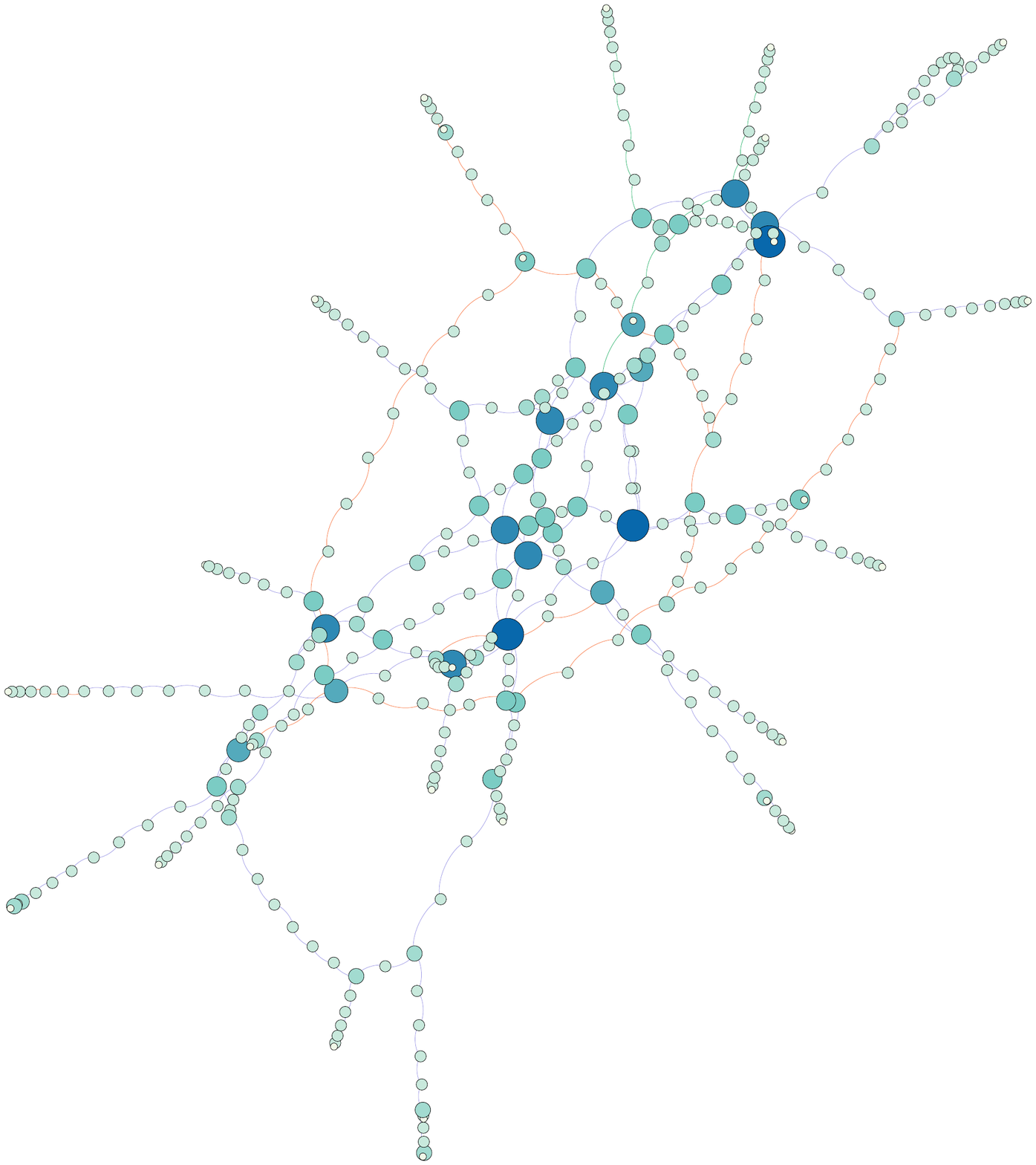}%
\label{fig:ltn}}
\hfil
\subfloat[Multilayer ForceAtlas Visualization of London Transportation Network]{\includegraphics[width=0.4\linewidth, trim={0.5cm 0.5cm 0.5cm 0.5}, clip]{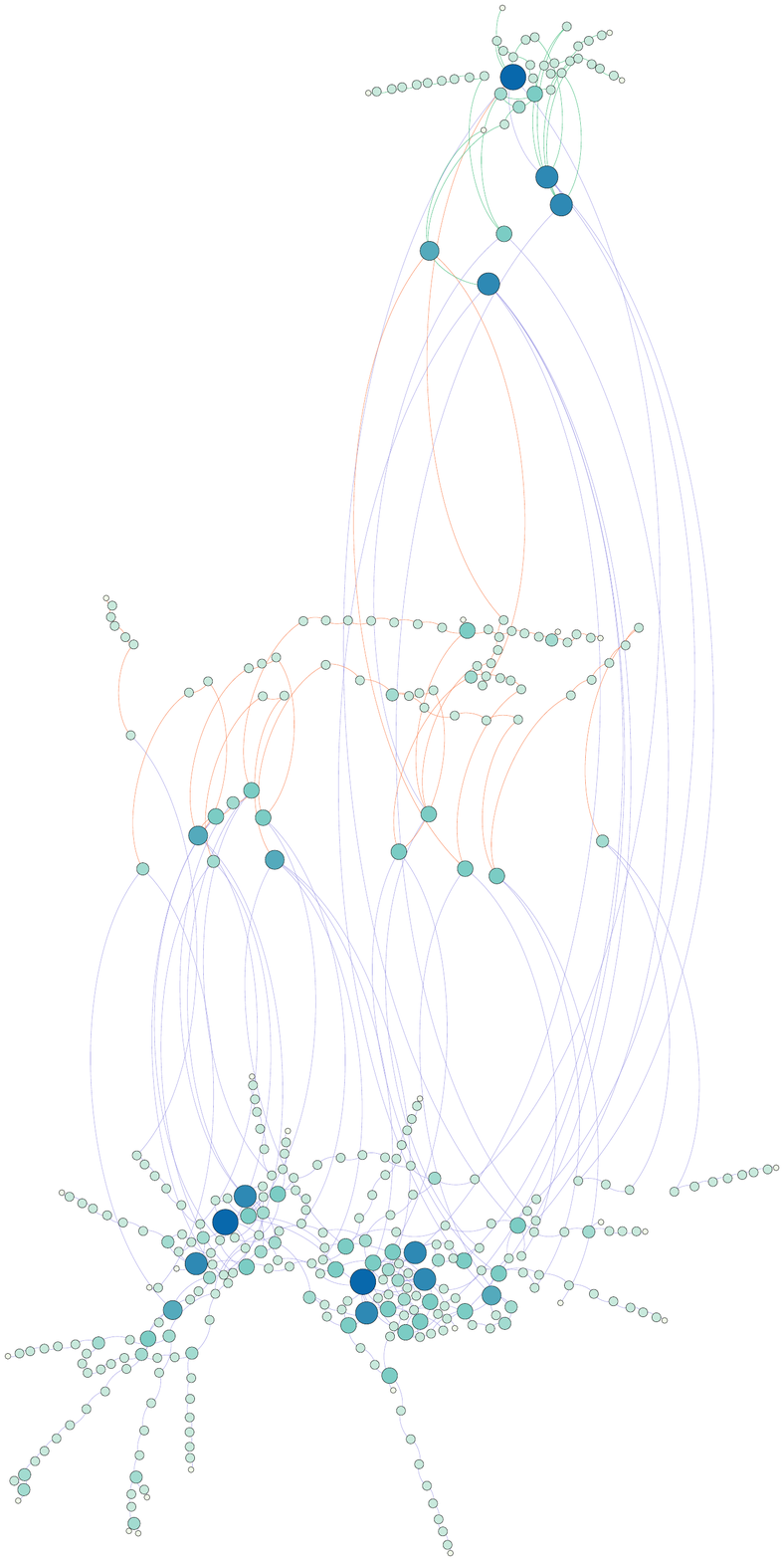}%
\label{fig:lfam}}\\
\subfloat[Multilayer ForceAtlas Visualization of London Transportation Network]{\includegraphics[width=0.4\linewidth, trim={0.5cm 0.5cm 0.5cm 0.5}, clip]{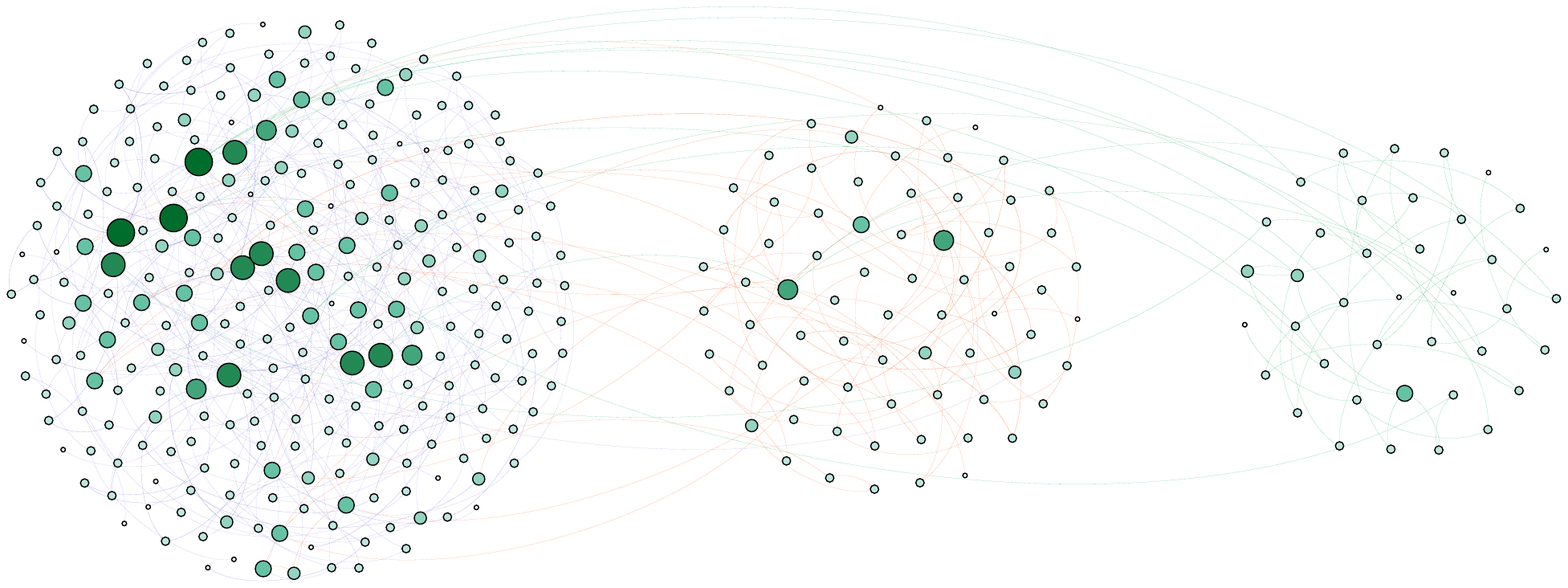}%
\label{fig:lfam2}}
\hfil
\subfloat[Multilayer Circular Layout Visualization of London Transportation Network]{\includegraphics[width=0.4\linewidth, trim={0.5cm 0.5cm 0.5cm 0.5}, clip]{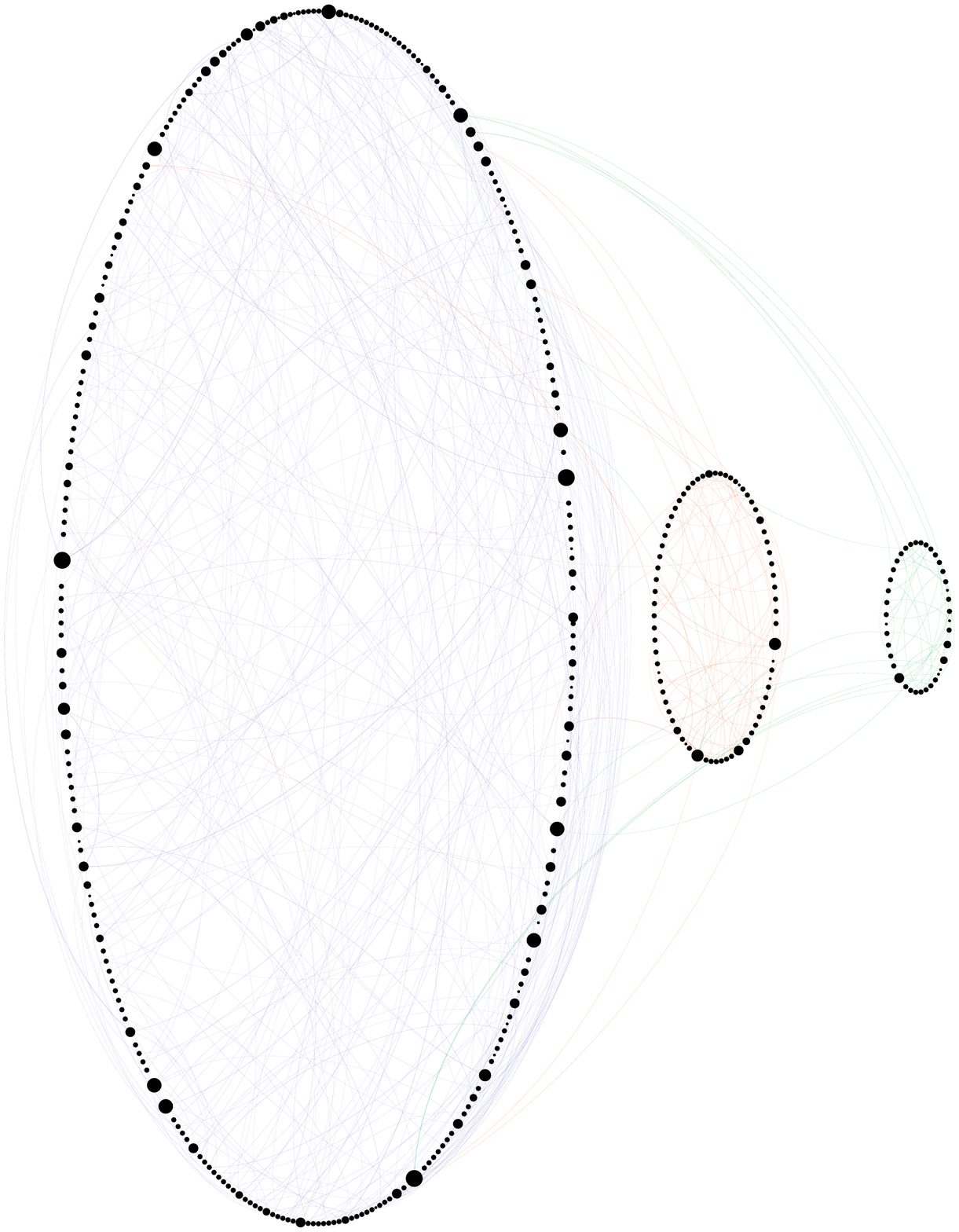}%
\label{fig:lfam3}}
\caption{Figures ~\ref{fig:ltn} and ~\ref{fig:lfam} shows London multiplex transportation network \cite{de2014navigability}, collected in 2013. Nodes are the train stations in London and edges are the existing route between stations. Stations on the underground, overground, and DLR (Docklands Light Railway) are considered in this network. This multiplex network makes use of three layers which are, the network of stations connected by the Tube, network of stations connected by Overground, and network of stations connected by DLR.  The node size indicates the number of routes pass through each station and edge color indicates type of transportation}
\label{fig:lfam_full}
\end{figure*}

\clearpage
\section{Conclusion and Future Work}
In the fields of social network analysis \cite{de2013mathematical} \cite{battiston2014structural}, economics, urban and international transportation\cite{cardillo2013emergence}\cite{gallotti2014anatomy}\cite{de2014navigability}, ecology\cite{stella2017parasite} \cite{pilosof2017multilayer} \cite{timoteo2018multilayer} \cite{costa2018species}, psychology\cite{fiori2007social}\cite{stella2017multiplex}, medicine, biology\cite{de2015structural}, commerce, climatology, physics\cite{de2016physics}, computational neuroscience \cite{timme2014multiplex} \cite{de2016mapping} \cite{battiston2017multilayer}, operations management, and finance, increasingly sophisticated attempts to model real-world systems as multidimensional networks have yielded valuable insight. By visualizing the complex networks in a multi-layer manner, users can further research into various properties of complex networks including multi-layer neighbours (all nodes connected to a node across layers), multi-layer path length \cite{magnani2013multidimensional}, dimension relevance (understand how important a particular layer is over the others for the connectivity of a node), network of layers \cite{de2013mathematical} (layers might be interconnected in such a way that their structure can be described by a network)
community discovery \cite{mucha2010community} \cite{de2015identifying} \cite{berlingerio2013abacus},clustering coefficients\cite{battiston2014structural} \cite{cozzo2015structure}
shortest path discovery \cite{brodka2011shortest}, information and disease spread through multi-layer networks\cite{granell2013dynamical} and so on.

By employing multi-dimensional/ multi-line multi-layer networks where each layer is of multiple dimensions or each layer has different sub-layers, more data can be visualized and researchers can take advantage of enhanced resolution in real-world data sets and exploit the current availability of big data to extract the ultimate and optimal representation of the underlying complex systems and mechanisms. 
%By introducing the multiviz plugin we hope to support the researchers, to enhance and continue the usage of Gephi, and with hope this plugin acts as a "point" to further introduction of new multilayer analysis tools to etc".
Many other visualization techniques like kamada kawai\cite{kamada1989algorithm}, Yifan Hu \cite{hu2005efficient}, OpenOrd\cite{martin2011openord}, GEM\cite{frick1994fast}  will be implemented in the plugin in future.

\appendix[ABBREVATIONS]
MLN - Multi-layer Networks\\
GDF - Graph Data Format\\
GML - Graph Modeling Language\\
DOT - GraphViz File Format\\
CSV - Comma Separated Files\\
TPL - Tulip Format\\
JSON - Javascript Object Notation\\
GEXF - Graph Exchange XML Format\\
SIF - Simple Interaction File Format\\
NNF - Nested Network Format\\
XGMML - Extensible Graph Markup and Modelling Language\\
SBML - Systems Biology Markup Language

\ifCLASSOPTIONcaptionsoff
  \newpage
\fi

\vskip 0pt plus -1fil
\begin{IEEEbiographynophoto}{Jayamohan Pillai}
is currently a master's student in Computer Science and Engineering of Amrita School of Computing, Amrita Vishwa Vidyapeetham, Amritapuri, India. He got his B.E in Computer Science and Engineering from Anna University, Chennai, India. His research interests are in network science, machine learning , learning analytics and artificial intelligence.
\end{IEEEbiographynophoto}
\vskip -2\baselineskip plus -1fil
\begin{IEEEbiographynophoto}{Ayan Chatterjee} is a doctoral student advised by Professor Tina Eliassi-Rad. His interests are in graph machine learning, specifically link prediction tasks. He got his B.E in Electronics and Telecommunication Engineering from Jadavpur University, India and a Master’s degree in Electronic Systems Engineering from the Indian Institute of Science. He worked with NVIDIA Graphics in developing GPU architectures for AI and Video Processing.
\end{IEEEbiographynophoto}
\vskip -2\baselineskip plus -1fil
\begin{IEEEbiographynophoto}{Geetha M.} received her B.Tech degree in Computer Science and Engineering from Kerala University, India in 2003, M.Tech degree in Computer Vision and Image Processing from Amrita University, Coimbatore India in 2005. She has been involved in teaching and research since 2006 and is currently with the faculty of the department of Computer Science and Engineering, Amrita Vishwa Vidyapeetham, Amrita University, Amritapuri, India. Her current research areas include computer vision, deep learning, video analytics, and brain network analysis. She is the receipient of a project funt from MeitY for providing Sign Language Accessibility to deaf community at the E-Governance Services of Govt. of India. She also has a patent published in the area of Sign Language Recognition. 
\end{IEEEbiographynophoto}
\vskip -2\baselineskip plus -1fil
\begin{IEEEbiographynophoto}{Amitava Mukherjee}
is a Professor at School of Computing, Amrita Vishwa Vidyapeetham, Amritapuri, India. Dr. Mukherjee is a Courtesy Adjunct Professor at College of Nanoscale Science and Engineering, SUNY Polytechnic Institute, Albany, NY since Aug, 2021. He got his Ph.D. in Computer Science and Engineering from Jadavpur University Kolkata, India. He has served as Dean at the School of Engineering and Technology and Head of the Department of Computer Science and Engineering at Adamas University, Barasat, Kolkata, India, as well as a visiting Professor on sabbatical at the University of New South Wales, Sydney and Royal Institute of Technology, Stockholm. He has extensive experience in realizing computational solutions to interdisciplinary fields of engineering, telecommunications, and biomedicine as well as global strategy implementation. He has worked as as senior manager of IBM India and principal consultant for PWC India.
\end{IEEEbiographynophoto}

\end{document}